\documentclass[fleqn,usenatbib]{mnras}

\usepackage{newtxtext,newtxmath}

\usepackage[T1]{fontenc}

\DeclareRobustCommand{\VAN}[3]{#2}
\let\VANthebibliography\thebibliography
\def\thebibliography{\DeclareRobustCommand{\VAN}[3]{##3}\VANthebibliography}

\usepackage{graphicx}	
\usepackage{amsmath}	

\title[Abell 2029/2033 filament]{A detailed study of the bridge of excess X-ray emission between the galaxy clusters Abell 2029 and Abell 2033}

\author[M. S. Mirakhor et al.]{
M. S. Mirakhor$^{1}$\thanks{E-mail: msm0033@uah.edu },
S. A. Walker$^{1}$, J. Runge$^{1}$
\\
$^{1}$Department of Physics and Astronomy, The University of Alabama in Huntsville, 301 Sparkman Drive, Huntsville, AL 35899, USA
}

\date{Accepted XXX. Received YYY; in original form ZZZ}

\pubyear{2015}

\begin{document}
\label{firstpage}
\pagerange{\pageref{firstpage}--\pageref{lastpage}}
\maketitle

\begin{abstract}
We examine \textit{Suzaku}, \textit{XMM--Newton}, and \textit{Chandra} observations of the Abell 2029/2033 system to investigate the nature of a bridge of X-ray emission joining the two galaxy clusters. By modelling the contributions from the outskirts of the two clusters, and excluding the emission from the southern infalling group and the background group LOS9, we find a significant excess of X-ray emission between the two clusters at the level of 6.5--7.0$\sigma$, depending on the choice of model, that cannot be explained by the overlap of the clusters. This excess component to the surface brightness is consistent with being emission from a filament with roughly 1.0 Mpc wide. The derived emission measure for the gas associated with the filament yields an average gas density of $3.7^{+1.0}_{-0.7} \times 10^{-5}$ cm$^{-3}$, corresponding roughly to 160 times the mean baryon density of the Universe. The \textit{Suzaku} X-ray spectrum of the excess emission indicates that it is significantly colder ($1.4_{-0.5}^{+0.7}$ keV) than the cluster outskirts emission from the two clusters ($\sim$ 5 keV), statistically consistent with the temperature expected from the hottest and densest parts of the warm-hot intergalactic medium (WHIM). The geometry, density, and temperature are similar to those found from X-ray studies of the Abell 222/223 filament.   
\end{abstract}



\section{Introduction}
\label{sec: intro}
Observations of the cosmic microwave background indicate that baryons account for about 5 per cent of the Universe’s total energy content \citep{ade2016planck}. However, the observed baryons in the local Universe is lower than the estimated baryons by about a factor of two \citep[e.g.][]{Fukugita1998,Bregman2007,Sinha2010}. Cosmological numerical simulations \citep[e.g.][]{Cen1999where,Dave2001,Dolag2006Simulating,Cui2019the,Martizzi2019Baryons,Tuominen2021eagle} suggest that the bulk of the missing baryons is in the form of a diffuse gas that resides in the cluster outskirts within filaments that connect clusters to the cosmic web. This diffuse gas, known as the warm-hot intergalactic medium (WHIM), is expected to have a density around 5--200 times the mean baryonic density of the Universe and a temperature in approximate range of $10^5 < T < 10^7$ K \citep[e.g.][]{Bregman2007}, making it difficult to detect with existing instruments. Understanding the composition, density, temperature and thermodynamic state of the filamentary gas is crucial for constraining models of large scale structure formation, and for accurate determinations of the baryon content of the Universe. 


Observationally, however, it is highly challenging to detect the WHIM due to its very low X-ray surface brightness \citep{Bregman2007}. One approach, which attempts to get around this limitation, is to search for the WHIM absorption lines along the line of sight to distant active galactic nuclei and quasars using high-resolution spectroscopy \citep[e.g.][]{Nicastro2005mass,Zappacosta2010studying,Nicastro2018observations}. Another approach to detect the WHIM is to observe large-scale filaments between a close pair of galaxy clusters, which are aligned such that the line of sight looks along the length of the filament, maximising the volume of gas we are looking through \citep[e.g.][]{Werner2008detection,Alvarez2018chandra}. Such filaments provide the best opportunity to trace the gas near the outskirts of cluster pairs, and to possibly detect the densest and hottest parts of the WHIM. 

Studying the filaments through their emission has the advantage that it allows us to spatially map all of their structure, whereas absorption studies rely on bright background objects, and are limited to a small number of sight lines through the filament. Detection of filaments through their X-ray emission remain rare and limited to a handful of cases, e.g., Abell 222/223 \citep{Werner2008detection}, Abell 399/401 \citep{Sakelliou2004xmm}, Abell 2804/2811 \citep{Sato2010study}, Abell 3556/3558 \citep{Mitsuishi2012search}, and Abell 3391/3395 \citep{Alvarez2018chandra,Reiprich2021the}.

\textit{ROSAT} and \textit{Suzaku} observations of the close pair of clusters Abell 2029 and Abell 2033 \citep{Walker2012_A2029} have shown there to be an excess in the X-ray emission in the direction connecting them. This system is a highly promising target for detecting and studying large-scale filamentary gas, given its geometry and angular separation. Abell 2029 is a massive galaxy cluster at redshift $z = 0.0767$ with a gas temperature of $kT = 8.5$ keV, situated in a small supercluster \citep{Einasto2001}, and the closest member of the supercluster to Abell 2029 is Abell 2033 ($z = 0.0818$, $kT = 4.2$ keV) at 37 arcmin to the north (corresponding to a distance of $\approx$ 3.4 Mpc). Assuming that the difference in redshift is due to the Hubble flow (so that there are no peculiar velocity effects), the line-of-sight distance between the two clusters is $\approx$ 20 Mpc, meaning that we are looking along the length of the filament, which will maximise its emission measure. This fortuitous geometry, therefore, enhances the possibility of detecting the WHIM. A similar geometry has been exploited for the system Abell 222/223 to detect the WHIM emission \citep{Werner2008detection}. 

In this work, we aim to investigate the filamentary structure of gas along the direction connecting the galaxy clusters Abell 2029 and Abell 2033 using \textit{XMM--Newton}, \textit{Chandra}, and \textit{Suzaku} observations. The rest of the paper is structured as follows: In Section \ref{sec: data}, we present the data reduction and analysis procedures; in Section \ref{sec: results}, we present our results; our findings are discussed in Section \ref{sec: discussion}; and our summary is presented in Section \ref{sec: summary}.

Throughout this work, we adopt a $\Lambda$ cold dark matter cosmology with $\Omega_{\rm{m}}=0.3$, $\Omega_{\rm{\Lambda}}=0.7$, and $H_0=100\,h_{100}$ km s$^{-1}$ Mpc$^{-1}$ with $h_{100}=0.7$. Uncertainties are at the 68 per cent confidence level, unless otherwise stated. At the redshift of the Abell 2029/2033 system ($z\approx0.08$), 1 arcmin corresponds roughly to 91 kpc. We follow the common practice of referring to $r_{200}$, the radius within which the mean density is 200 times the critical density of the Universe, as the virial radius.

\begin{figure*}

\begin{center}

\includegraphics[width=\textwidth]{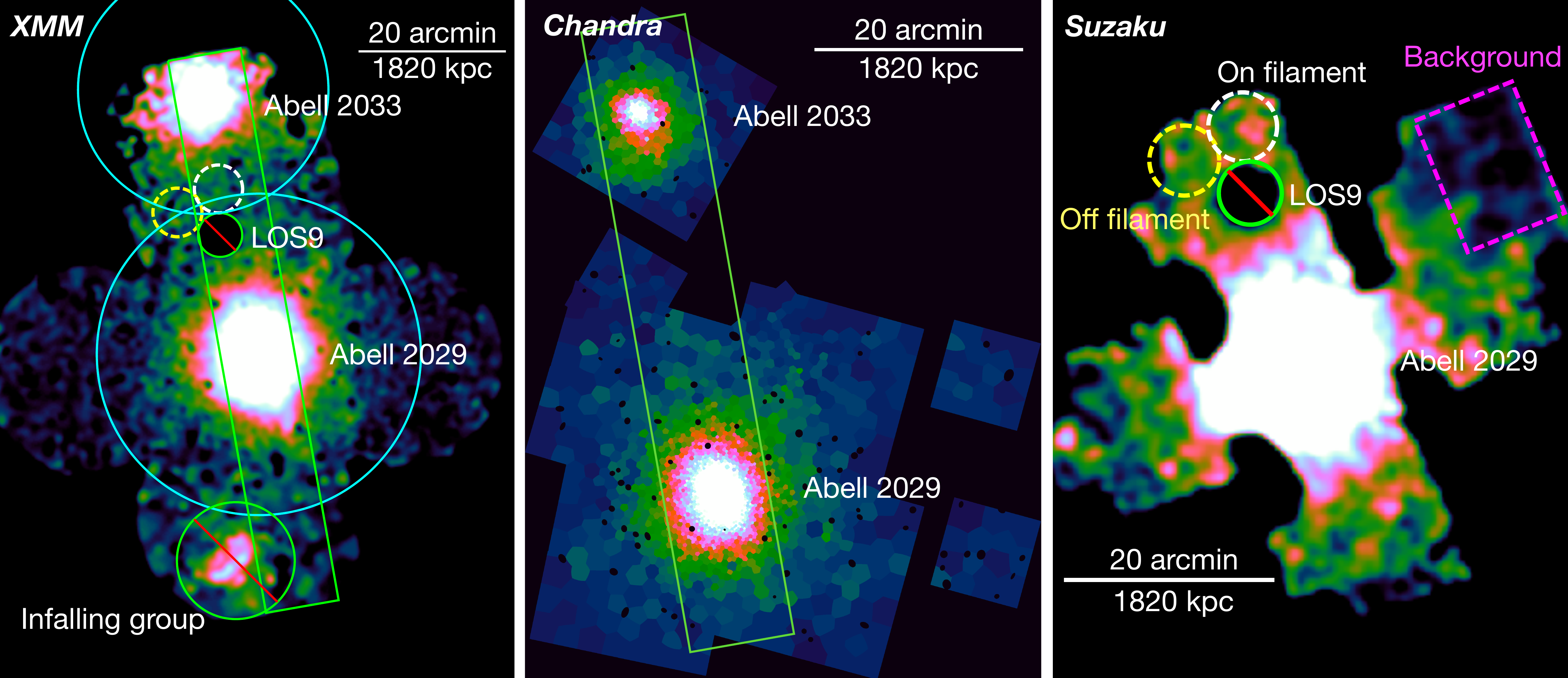} 
\end{center}
\caption{\textit{Left:} \textit{XMM--Newton} mosaicked and Voronoi tessellated image of the Abell 2029/2033 system in the 0.7--1.2 keV energy band. The image is smoothed by a Gaussian kernel with $\sigma=10$ arcsec to highlight the filamentary gas between the two clusters. The virial radii of the two clusters \citep[22 arcmin for Abell 2029 and 17 arcmin for Abell 2033;][]{Walker2012_A2029} are shown as the cyan circles. \textit{Middle:} \textit{Chandra} mosaicked and Voronoi tessellated image of the Abell 2029/2033 system, mainly used to model the shape of the gas distribution to the north of Abell 2033. \textit{Right:} \textit{Suzaku} mosaic of the Abell 2029/2033 system. The green boxes show the regions in which the X-ray counts were extracted and used to estimate the surface brightness profile for the Abell 2029/2033 system. We excluded the emission from the southern infalling group and the background group LOS9 (the green circles in the left-hand panel). The regions used to extract the \textit{Suzaku} spectra for the filament are indicated by white and yellow dashed circles. The magenta dashed box shows the region in which the background spectrum was extracted.}
\label{fig: A2029/2033}
\vspace{-0.2cm}
\end{figure*}

\section{Data and analysis procedure}
\label{sec: data}
\subsection{\textit{XMM--Newton}}
In this work, we used a total of 7 \textit{XMM--Newton} observations, covering the Abell 2029/2033 field and the gaseous bridge between the two clusters. The observations were carried out in the period between 2008 and 2019. The exposure time ranges between 31 and 50 ks each, for a total exposure time of around 282 ks. A summary of the observations is given in Table \ref{table: data}. We reduced the data using the \textit{XMM--Newton} Science Analysis System (\textit{XMM}-\textsc{sas}) version 19.1 and current calibration files (CCF), following the procedure illustrated in the Extended Source Analysis Software (\textsc{esas}) cookbook\footnote{https://heasarc.gsfc.nasa.gov/docs/xmm/esas/cookbook/xmm-esas.html}. Below, we provide a summary of the data reduction procedure carried out on the \textit{XMM--Newton} observations. 

\begin{table*}
\begin{minipage}{130mm}
    \centering
    \caption{Summary of the \textit{XMM--Newton}, \textit{Chandra}, and \textit{Suzaku} observations for the Abell 2029/2033 field}
    \begin{tabular}{cccccccc}
    \hline
    Observatory   &   Obs. ID   & RA     & Dec.    & Obs. Date  & \multicolumn{3}{c}{Filtered Exposure Time (ks)}\\
    \cline{6-8}
                  &             &        &         &            & MOS1/MOS2/PN  & XIS & ACIS-I  \\
        \hline
\textit{XMM--Newton} & 0551780201   & 15 10 58.72 & +05 45 42.1 & 2008-07-17  & 33.1/34.0/15.1 & --- & ---  \\
\textit{XMM--Newton} & 0551780301   & 15 10 58.72 & +05 45 42.1 & 2008-07-19  & 39.0/40.9/27.4 & --- & ---  \\
\textit{XMM--Newton} & 0744410901   & 15 11 09.74 & +06 04 23.1 & 2015-02-08  & 17.5/22.3 /6.6 & --- & ---   \\
\textit{XMM--Newton} & 0744411001   & 15 09 39.59 & +05 44 12.0 & 2015-07-27  & 42.6/43.3/39.2 & --- & ---  \\
\textit{XMM--Newton} & 0744411201   & 15 12 28.85 & +05 42 27.3 & 2015-01-31  & 26.0/26.7/20.6  &--- & ---  \\
\textit{XMM--Newton} & 0744411101   & 15 11 05.72 & +05 22 32.0 & 2015-02-22  & 11.0/16.6/4.9 & --- & ---  \\
\textit{XMM--Newton} & 0823650101   & 15 11 14.54 & +06 12 20.9 & 2018-08-21  & 14.4/16.1/4.5  & --- & ---  \\
\textit{Suzaku}      & 804024010    & 15 10 57.82 & +05 44 59.3 & 2010-01-28  & --- & 14.2 & ---   \\ 
\textit{Suzaku}      & 804024020    & 15 11 24.65 & +06 00 38.9 & 2010-01-28  & --- & 43.4 & --- \\ 
\textit{Suzaku}      & 804024030    & 15 10 31.46 & +05 29 14.6 & 2010-01-28  & ---  & 33.4 & ---  \\ 
\textit{Suzaku}      & 804024040    & 15 12 01.27 & +05 38 58.6 & 2010-01-29  & --- & 40.4 & --- \\ 
\textit{Suzaku}      & 804024050    & 15 09 55.39 & +05 28 50.9 & 2010-01-30  & --- & 36.0 & --- \\ 
\textit{Suzaku}      & 802060010    & 15 09 51.46 & +06 01 25.7 & 2008-01-08  & --- & 289.2 & --- \\ 
\textit{Chandra}     & 6101         & 15 10 56.20 &	+05 44 40.5 & 2004-12-17  & --- & --- & 9.9  \\ 
\textit{Chandra}     & 10434        & 15 10 18.33 &	+05 53 47.7 & 2009-04-01  & --- & --- & 5.1    \\
\textit{Chandra}     & 10435        & 15 11 26.37 & +05 53 47.7 & 2009-04-01  & --- & --- & 4.7    \\
\textit{Chandra}     & 10436        & 15 10 18.35 & +05 36 52.4 & 2009-04-01  & --- & --- & 4.7    \\
\textit{Chandra}     & 10437        & 15 11 26.36 & +05 36 52.4 & 2009-04-01  & --- & --- & 4.7    \\
\textit{Chandra}     & 15167        & 15 11 23.50 & +06 20 08.4 & 2013-05-22  & --- & --- & 9.0    \\
  \hline
    \end{tabular}
    \label{table: data}
\end{minipage}
\end{table*}

We initiated the data processing by running the \textit{epchain} and \textit{emchain} tasks, followed by the tasks \textit{mos-filter} and \textit{pn-filter} to remove soft proton flares and create clean event files for MOS and PN detectors. The data of the MOS detectors were screened for CCDs in anomalous states, and any affected CCDs were excluded from further analysis. We also run the \textit{mos-spectra} and \textit{pn-spectra} scripts to create spectra, response matrix files (RMFs), ancillary response files (ARFs), and exposure maps for the entire region of interest. These files were then turned into the quiescent particle background spectra and images in the MOS and PN coordinates by running the \textit{mos-back} and \textit{pn-back} scripts. The data were further examined for any soft proton contamination that may have remained after the initial light-curve screening by running the \textit{proton} task to create images of the soft proton contamination. The \textit{XMM} background was subtracted using closed filter observations. 

The analysis procedure described above was applied to each \textit{XMM--Newton} observation listed in Table \ref{table: data}. After weighting the \textit{XMM--Newton} detectors by their effective area, the primary components required for a background-subtracted and exposure-corrected image from all observations and three detectors were merged and adaptively smoothed into a single image. The left-hand panel of Fig. \ref{fig: A2029/2033} shows a background-subtracted and exposure-corrected mosaicked image of the Abell 2029 and Abell 2033 system in the energy band 0.7--1.2 keV. We applied a Voronoi tessellation algorithm \citep{diehl2006} on the mosaicked count image to create an adaptively binned image with each tessellated region has at least 20 counts. Point sources and extended substructures in the field of Abell 2029 and Abell 2033 were detected and masked by running the \textsc{esas} source-detection tool \textit{cheese}. The image was also examined for any remaining point sources that were missed using \textit{cheese} by running the \textit{Chandra} source-detection tool \textit{wavdetect} with wavelet scales of 2, 4, 8, 16, and 32 pixels.

\subsection{\textit{Chandra}}
Abell 2029 was observed with \textit{Chandra} several times during the period between 2004 and 2010, whereas Abell 2033 was observed once by \textit{Chandra} in 2013. Table \ref{table: data} gives a summary of the \textit{Chandra} pointings used in this work. We used \textit{Chandra} Interactive Analysis of Observations (\textsc{ciao}) version 4.13, with the latest calibration database (\textsc{caldb}) version 4.9.5 to reduce the data. We reprocessed the data by running the \textit{chandra\_repro} script to carry out the recommended data preparation such as checking the source coordinate, removing streak events, filtering the event file to good time intervals, and identifying the bad pixels. This script creates a new level $=2$ event file, and a new bad pixel file. We then run the \textit{merge\_obs} script to create merged event files, exposure maps, and exposure-corrected images. The background counts were extracted from a circular region to the west of Abell 2029 at a radius beyond $r_{200}$ of the cluster. Point sources and extended substructures in the Abell 2029/2033 field were detected and removed using the \textit{Chandra} source-detection tool \textit{wavdetect} with wavelet scales of 2, 4, 8, 16, and 32 pixels. For our background subtraction, we used stowed backgrounds, following the method used in \citet{Wang2014}.

Similar to the \textit{XMM--Newton} data, we applied a Voronoi tessellation algorithm on the mosaicked count image to create an adaptively binned image with each tessellated region has at least 20 counts. In the right-panel of Fig. \ref{fig: A2029/2033}, we present the \textit{Chandra} mosaicked and Voronoi tessellated image of the Abell 2029/2033 system. At present, the gaseous bridge between the two clusters has not been observed by \textit{Chandra}. However, the \textit{Chandra} data provide a better azimuthal coverage to the north of Abell 2033 than \textit{XMM--Newton}, allowing us to accurately model the shape of the X-ray emission in Abell 2033 and subtract its contribution from the gaseous bridge connecting the two clusters.

\subsection{\textit{Suzaku}}

To reduce and analyze the \textit{Suzaku} data we follow the methods described in \cite{Walker2012_A2029} and \cite{Walker2013_Centaurus}. The \textit{Suzaku} observations (tabulated in Table \ref{table: data}) were reprocessed using \textsc{aepipeline}, and the standard cleaning outlined in the \textit{Suzaku} ABC guide\footnote{https://heasarc.gsfc.nasa.gov/docs/suzaku/analysis/abc/}. 

Images were extracted in the 0.5--2.0 keV band, exposure maps were produced using \textsc{xissim}, and non-X-ray background (NXB) images were created using \textsc{xisnxbgen}. A background subtracted, exposure corrected mosaic of the \textit{Suzaku} observations was created using all of the available detectors (XIS0, XIS1 and XIS3) and is shown in the right-hand panel of Fig. \ref{fig: A2029/2033}. As described in \cite{Walker2012_A2029}, the measurements from the Solar Wind Ion Composition Exlorer (SWICS) and the Advanced Composition Explorer (ACE) taken during the \textit{Suzaku} observations of Abell 2029 found that the ratio of O$^{7+}$ to O$^{6+}$ in the solar wind was below the threshold of 0.2, which means the effects of solar wind charge exchange are negligible \citep{Snowden2004}.

\begin{figure*}
\begin{center}
\includegraphics[width=0.9\textwidth]{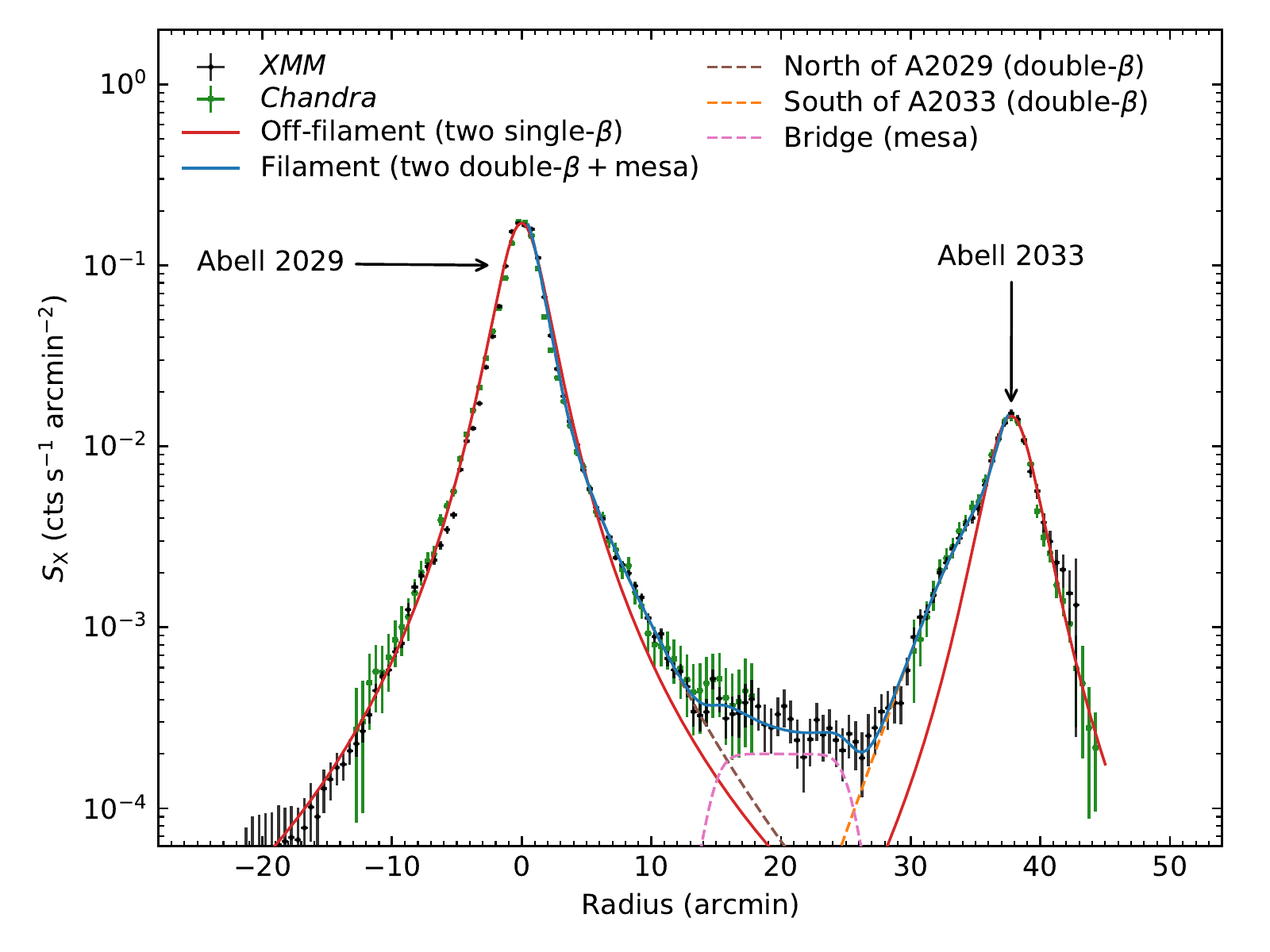} 
\end{center}
\vspace{-0.5cm}
\caption{Background-subtracted surface brightness profile for the Abell 2029/2033 field using the \textit{XMM--Newton} and \textit{Chandra} observations, along with two models for the cluster contribution to the emission. The Chandra data have been appropriately scaled to account for the difference in effective area between XMM and Chandra, and emission from background groups has been masked out. The \textit{x}-axis shows the distance from the centre of Abell 2029, going northwards towards Abell 2033. As clearly suggested by the observed data, there is a significant excess in the direction connecting the two clusters, near their virial radii between 15--25 arcmin to the north of Abell 2029. The red solid line is the best single $\beta$-model fit to the parts of the clusters directed away from the filament direction, excluding the emission from the southern infalling group. This model accurately reproduces the X-ray data to the south of Abell 2029 and the north of Abell 2033 over most of the radial range. However, the model significantly underestimates the true emission observed along the filament direction, extending from the north of Abell 2029 at around $+5$ arcmin out to the south of Abell 2033 at around $+35$ arcmin. The blue solid line is the best model fit to the surface brightness profile along the filament direction, made up of two double-$\beta$ components for each cluster and a "mesa-like" function. The components of this model are shown as dashed lines. The model provides a good fit to the X-ray surface brightness profile over most of the radial range along the bridge region. }
\label{fig: sb_filament}
\end{figure*}

\section{Results}
\label{sec: results}
\subsection{Surface brightness}
\label{sec: SB}
As mentioned in Section \ref{sec: intro}, Abell 2029 is a massive cluster in the local Universe, and Abell 2033 is less massive than Abell 2029 at slightly higher redshift, located at a distance of 37 arcmin (corresponding roughly to a distance of 3.4 Mpc) to the north. There is an extended bridge of hot gas connecting the two, and an infalling group located to the south of Abell 2029 (see Fig. \ref{fig: A2029/2033}). There is also a small background group \citep[LOS9;][]{Sohn2019the} coincident with the bridge, residing outside the Abell 2029/2033 system at $z \sim 0.223$. 

Using the \textit{XMM--Newton} and \textit{Chandra} data, we derived the surface brightness profile for the Abell 2029/2033 system by extracting counts in annuli with equally spaced bins of 30 arcsec, starting from the south of Abell 2029 and going upwards to the north of Abell 2033 with an inclination angle of 102 degree with respect to the right ascension axis. We excluded the emission from the southern infalling group and the background group LOS9 (see the next section, \ref{sec: LOS9}). X-ray background counts were extracted from a circular region to the west of Abell 2029 at a radius beyond $r_{200}$ of the cluster. In Fig. \ref{fig: sb_filament}, we show the background-subtracted, exposure-corrected surface brightness profile for the Abell 2029/2033 field using the \textit{XMM--Newton} and \textit{Chandra} observations. The \textit{Chandra} surface brightness has been appropriately scaled to account for the difference in the effecitve area between \textit{XMM} and \textit{Chandra}. The \textit{Chandra} data provide greater coverage of Abell 2033 to the north than the \textit{XMM} data. A significant excess emission is apparent in the direction connecting the two clusters near their virial radii, and the surface brightness profile in this region is reasonably flat. 


\begin{table*}
\begin{minipage}{178mm}
    \centering
    \caption{Best-fitting parameters for the models used in this work.}
    \begin{tabular}{ccccccc}
    \hline
       &    & \multicolumn{5}{c}{Abell 2029} \\
    \cline{3-7}
    Region   & Model & $S_{\rm{X01}}$  &  $r_{\rm{c1}}$  & $\beta$  & $S_{\rm{X02}}$   & $r_{\rm{c2}}$   \\
             &    & ($10^{-2}$ cts s$^{-1}$ arcmin$^{-2}$)   & (arcmin) &    &  ($10^{-2}$ cts s$^{-1}$ arcmin$^{-2}$)  &   (arcmin)                  \\           
        \hline
Off-filament & Single-$\beta$ & $17.11_{-0.90}^{+0.97}$ & $2.33_{-0.10}^{+0.11}$ & $0.79_{-0.02}^{+0.02}$ & --- & ---  \\ 
Filament  & Double-$\beta$ & $19.14_{-0.85}^{+0.94}$ & $1.71_{-0.11}^{+0.12}$ & $0.69_{-0.03}^{+0.03}$ & $0.15_{-0.03}^{+0.04}$ & $7.65_{-1.71}^{+1.90}$  \\ 
Filament  & Double-$\beta$ + single-$\beta$ (bridge) & $17.27_{-0.65}^{+0.68}$ & $2.68_{-0.27}^{+0.33}$ & $1.18_{-0.14}^{+0.19}$ & $1.11_{-0.14}^{+0.15}$ & $8.97_{-0.88}^{+1.00}$   \\ 
Filament  & Double-$\beta$ + mesa (bridge) & $17.55_{-0.58}^{+0.60}$ & $2.49_{-0.20}^{+0.22}$ & $1.07_{-0.10}^{+0.11}$ & $0.95_{-0.10}^{+0.10}$ & $8.66_{-0.73}^{+0.76}$   \\ 
  \hline
        &    & \multicolumn{5}{c}{Abell 2033} \\
    \cline{3-7}
    Region   & Model  & $S_{\rm{X01}}$  & $r_{\rm{c1}}$  & $\beta$  & $S_{\rm{X02}}$   & $r_{\rm{c2}}$   \\
             &    & ($10^{-2}$ cts s$^{-1}$ arcmin$^{-2}$)   & (arcmin) &   &  ($10^{-2}$ cts s$^{-1}$ arcmin$^{-2}$)  &   (arcmin)                  \\           
        \hline
Off-filament & Single-$\beta$ & $1.47_{-0.09}^{+0.10}$ & $2.67_{-0.26}^{+0.31}$ & $0.86_{-0.05}^{+0.06}$ & --- & --- \\ 
Filament  & Double-$\beta$  & $0.96_{-0.09}^{+0.12}$ & $2.35_{-0.29}^{+0.38}$ & $1.33_{-0.11}^{+0.20}$ & $0.56_{-0.05}^{+0.05}$ & $8.71_{-0.52}^{+1.01}$   \\ 
Filament  & Double-$\beta$ + single-$\beta$ (bridge) & $0.92_{-0.11}^{+0.12}$ & $2.67_{-0.35}^{+0.35}$ & $1.80_{-0.22}^{+0.14}$ & $0.60_{-0.04}^{+0.04}$ & $10.32_{-0.94}^{+0.67}$   \\ 
Filament  & Double-$\beta$ + mesa (bridge) & $0.93_{-0.07}^{+0.07}$ & $2.72_{-0.34}^{+0.33}$ & $1.79_{-0.22}^{+0.15}$ & $0.59_{-0.04}^{+0.04}$ & $10.55_{-0.97}^{+0.69}$  \\ 
  \hline
         &    & & \multicolumn{3}{c}{Bridge} & \\
    \cline{4-6}
    Region   & Model &  & $S_{\rm{fil}}$  &  $\alpha_{\rm{fil}}$ & $r_{\rm{c,fil}}$ &   \\
             &       &  & ($10^{-3}$ cts s$^{-1}$ arcmin$^{-2}$) & (arcmin) & (arcmin) &    \\           
        \hline
        
Filament  & Two double-$\beta$ + single-$\beta$ & & $0.25_{-0.03}^{+0.03}$ & $19.47_{-0.81}^{+0.76}$ & $12.10_{-2.13}^{+3.35}$ &  \\
Filament  & Two double-$\beta$ + mesa & & $0.20_{-0.02}^{+0.02}$ & $20.01_{-0.37}^{+0.39}$ & $5.57_{-0.38}^{+0.39}$ &  \\
  \hline
    \end{tabular}
    \label{table: best_fitting_parameters}
\end{minipage}
\end{table*}


In order to measure the emission coming just from the gaseous bridge that connecting the two clusters, we need to model the contributions to the emission coming from the galaxy clusters Abell 2029 and Abell 2033, and subtract their contributions from the bridge. For this purpose, we first used a single $\beta$-model \citep{Cavaliere1976} to determine the shapes of the X-ray emission in Abell 2029 and Abell 2033. The surface brightness profile of the single $\beta$-model is expressed by  
\begin{equation}
    S_{\rm{X}}(r)=S_{\rm{X0}}\bigg[1+\bigg(\frac{r}{r_{\rm{c}}}\bigg)^2\bigg]^{-3\beta+1/2}, 
    \label{eq: single_beta_model}
\end{equation}
where $S_{\rm{X0}}$ is the central surface brightness parameter, $\beta$ describes the shape of the gas distribution, and $r_{\rm{c}}$ is the core radius.

For the fitting procedure, we used the affine invariant Markov chain Monte Carlo ensemble sampler implemented in the \textsc{emcee} package \citep{goodman2010ensemble}. By excluding the emission from the southern infalling group and the background LOS9 group, we simultaneously fitted the gas distribution to the south of Abell 2029 to a single $\beta$-model, and the gas distribution to the north of Abell 2033 to another single $\beta$-model, i.e. using two single $\beta$-models along the off-filament direction. For both clusters, we let the parameters in equation (\ref{eq: single_beta_model}) free to vary, and fix the cluster centre at the X-ray peak. By doing that, we find that the two-single $\beta$-model accurately reproduces the X-ray surface brightness profiles to the south of Abell 2029 and the north of Abell 2033  (i.e. the directions away from the possible filament) over most of the radial range (see the red solid line in Fig. \ref{fig: sb_filament}). We found the reduced $\chi^2$ associated with this fit is 1.03 for 86 degrees of freedom. The values of the best-fitting parameters for this fit are listed in Table \ref{table: best_fitting_parameters}. This model, however, significantly underestimates the surface brightness profile derived along the direction connecting the two clusters, extending from the north of Abell 2029 at around $+5$ arcmin out to the south of Abell 2033 at around $+35$ arcmin from the Abell 2029 centre.

We then repeated the fitting process, but this time by modelling the surface brightness profile along the filament direction to two single $\beta$-models. In this case, a single $\beta$-model is used to describe the gas distribution to the north of Abell 2029 and another one is used to describe the gas distribution to the south of Abell 2033. As for the off-filament case, we let the parameters in equation (\ref{eq: single_beta_model}) for both clusters free to vary, and fix the cluster centre at the X-ray peak. We find that the model does not provide a good fit to the surface brightness profile derived for the north of Abell 2029 and the south of Abell 2033 over most of the radial range. The value of the reduced $\chi^2$ associated with this fit is 1.88. 

However, we find that a slightly better fit for the X-ray surface brightness along the filament direction can be obtained by simultaneously fitting the north of Abell 2029 to a double $\beta$-model \citep{Mohr1999properties} and the south of Abell 2033 to another double $\beta$-model, i.e. using two double $\beta$-models along the filament direction. The surface brightness profile of the double $\beta$-model is expressed by  
\begin{equation}
    S_{\rm{X}}(r)=S_{\rm{X01}}\bigg[1+\bigg(\frac{r}{r_{\rm{c1}}}\bigg)^2\bigg]^{-3\beta+1/2} + S_{\rm{X02}}\bigg[1+\bigg(\frac{r}{r_{\rm{c2}}}\bigg)^2\bigg]^{-3\beta+1/2},
    \label{eq: double_beta_model}
\end{equation}
where we used the same $\beta$ for both components to reduce the degeneracy between the shape parameters. Here, the first component in equation (\ref{eq: double_beta_model}) is used to describe the narrow, sharply-peaked central surface brightness, whereas the second component is used to describe the broad, shallow outer surface brightness profile.  

Although the model provides a reasonable fit to the surface brightness profile at small to intermediate radii, it does not fit well the observed surface brightness profile for the bridge region. The value of the reduced $\chi^2$ associated with this fit is 1.40 for 116 degrees of freedom, implying that the fit can be rejected at the 99.71 per cent confidence level.

Using the Sunyaev--Zel'dovich (SZ) effect data from the Atacama Cosmology Telescope and \textit{Planck} satellite, \citet{Hincks2021high} recently studied the gas in the Abell 399/401 field, a close pair of galaxy clusters, and they used two ad hoc models to fit the apparent flatness of the SZ signal in the bridge region. In the first case, they used a single $\beta$-model, but with $\beta$ fixed at 4/3, which is suitable for an isothermal cylinder in hydrostatic equilibrium \citep{Ostriker1964the}. In the second case, they fitted the excess SZ signal in the bridge region to a flat, "mesa-like" function. The model, which is named as "mesa", is not physically motivated, but is rather an ad hoc model tailored to the observed data with minimal assumptions about its shape. \citet{Hincks2021high} have done their analysis on the 2D SZ image, but here we carry out our analysis on the 1D surface brightness profile. Hence, the 1D mesa model of the surface brightness is of the form 
\begin{equation}
    m (r)=S_{\rm{fil}}\bigg[1 + \bigg(\frac{r-\alpha_{\rm{fil}}}{r_{\rm{c,fil}}}\bigg)^{8}\bigg]^{-1}, 
    \label{eq: mesa}
\end{equation}
where $S_{\rm{fil}}$ is the surface brightness of the mesa, $\alpha_{\rm{fil}}$ is the mesa centre, and $r_{\rm{c,fil}}$ is the characteristic radius.

In this work, we adopt both cases to fit for the surface brightness in the bridge region. Along the filament direction, therefore, we use two double-$\beta$ components, one for each cluster, and with a single-$\beta$ component or a "mesa-like" function (equation \ref{eq: mesa}) for the bridge region. The model, in both cases, has sufficient degrees of freedom to accurately fit the X-ray surface brightness profile along the filament direction, with ten parameters to model the surface brightness distributions to the north of the Abell 2029 and to the south of Abell 2033 and three additional parameters to model the excess emission in the bridge region. We find that both models provide a good fit to the surface brightness over most of the radial range, including the bridge region. Similar to \citet{Hincks2021high}, a large core radius ($r_c=12.10_{-2.13}^{+3.35}$ arcmin, corresponding roughly to $1.10_{-0.19}^{+0.31}$ Mpc at the redshift of the Abell 2029/2033 system) is obtained for the bridge when it fitted with a single $\beta$-model. Values of the best-fitting parameters associated with these fits are reported in Table \ref{table: best_fitting_parameters}. The reduced $\chi^2$ associated with the models that include a "mesa-like" function and a single $\beta$-model for the bridge are 0.91 and 0.97 with 113 degrees of freedom, respectively. This suggests that the models that include a bridge component provide a better fit than the previous model (two double-$\beta$ components for the clusters with no component for the bridge).

\begin{figure}

\begin{center}

\includegraphics[width=1.0\columnwidth]{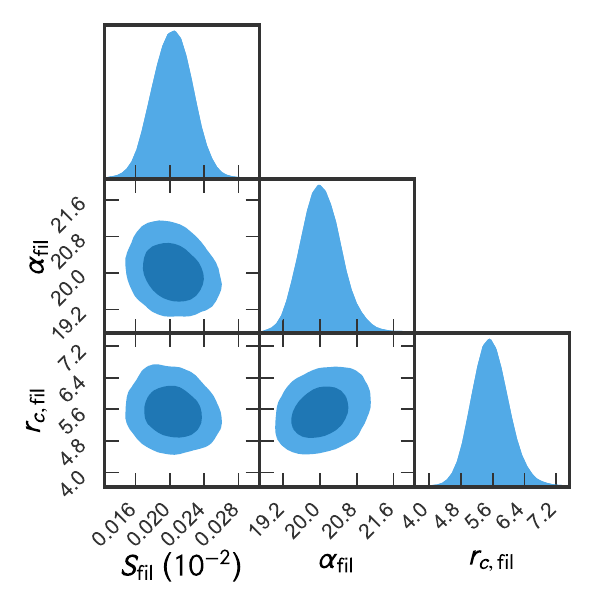} 
\end{center}
\vspace{-0.5cm}
\caption{Posterior distributions for the mesa parameters obtained from the model that consists of two double-$\beta$ components plus a "mesa-like" function. There is a weak degeneracy between the fitting parameters. Contour levels refer to the 68th and 95th percentiles. The $S_{\rm{fil}}$ parameter is in units of cts s$^{-1}$ arcmin$^{-2}$, $\alpha_{\rm{fil}}$ and $r_{\rm{c, fil}}$ are in units of arcmin.}
\label{fig: posterior for mesa}
\end{figure}

The blue solid line in Fig. \ref{fig: sb_filament} is the best-fitting model (two double-$\beta$ components plus a "mesa-like" function) to the surface brightness profile along the filament direction. The model accurately reproduces the X-ray data over most of the radial range, including the bridge region. Fig. \ref{fig: posterior for mesa} shows the posterior distributions for the mesa parameters obtained from the model that consists of two double-$\beta$ components plus a "mesa-like" function. The width of this "mesa" function is $2\times 5.57$ arcmin $=11.1$ arcmin $=1.0$ Mpc, which is similar to the width of the filament seen in Abell 222/223 in \citet{Werner2008detection}.

\begin{figure}

\begin{center}

\includegraphics[width=1.0\columnwidth]{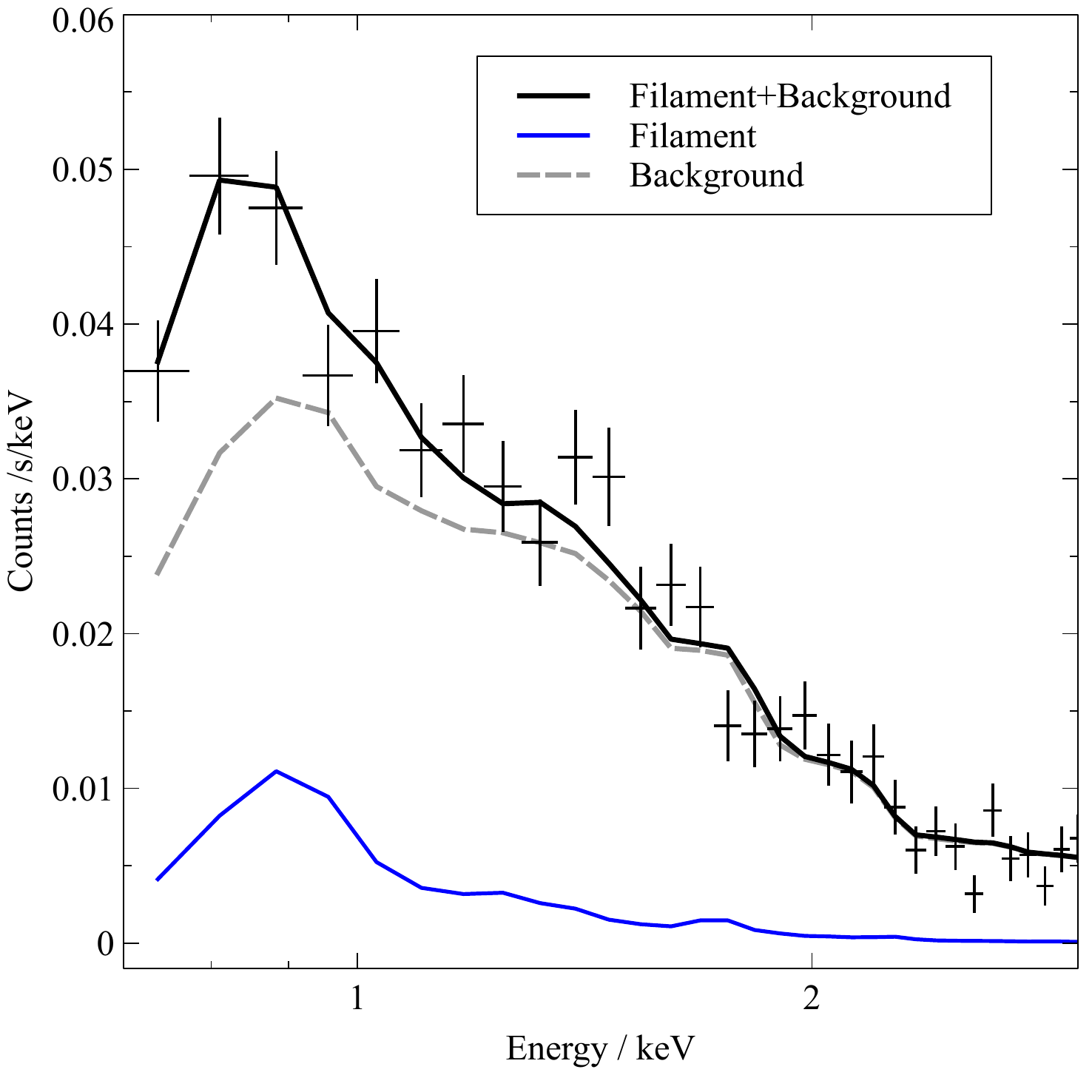} 
\end{center}
\vspace{-0.5cm}
\caption{\textit{Suzaku} spectral fit to the filament region marked in the right hand panel of Fig. \ref{fig: A2029/2033}. The blue line shows the excess emission over the background level (dashed grey) which is the sum of the X-ray background and the emission from the outskirts of Abell 2029/2033.}
\label{fig: SuzakuSpectrum}
\end{figure}



 \subsection{Background group-LOS9}
\label{sec: LOS9}
\citet{Sohn2019the} studied the structure of the Abell 2029/2033 system using spectroscopy along with X-ray and weak lensing data, and identified at least 12 foreground and background galaxy groups in the Abell 2029/2033 field. The location of one of these groups, LOS9, coincides with the location of the bridge, and resides outside the Abell 2029/2033 field at higher redshift ($z \sim 0.223$). The LOS9 system has associated X-ray emission with the flux of $2 \times 10^{-13}$ erg s$^{-1}$ cm$^{-2}$, which corresponds to an X-ray luminosity of $3.2 \times 10^{43}$ erg s$^{-1}$ at the redshift of the group. Since the data allows us to resolve point sources down to a threshold flux of around $4.0 \times 10^{-15}$ erg cm$^{-2}$ s$^{-1}$ in the 0.5--2.0 keV energy band, this group is well resolved in our \textit{XMM} image. Previous studies of this system in X-rays \citep{Sohn2019the} have used only the \textit{ROSAT} PSPC observation, in which the LOS9 group is significantly off-axis (17.5 arcmin) and significantly broadened by \textit{ROSAT}'s large off-axis PSF.


\begin{figure}

\begin{center}

\includegraphics[width=1.0\columnwidth]{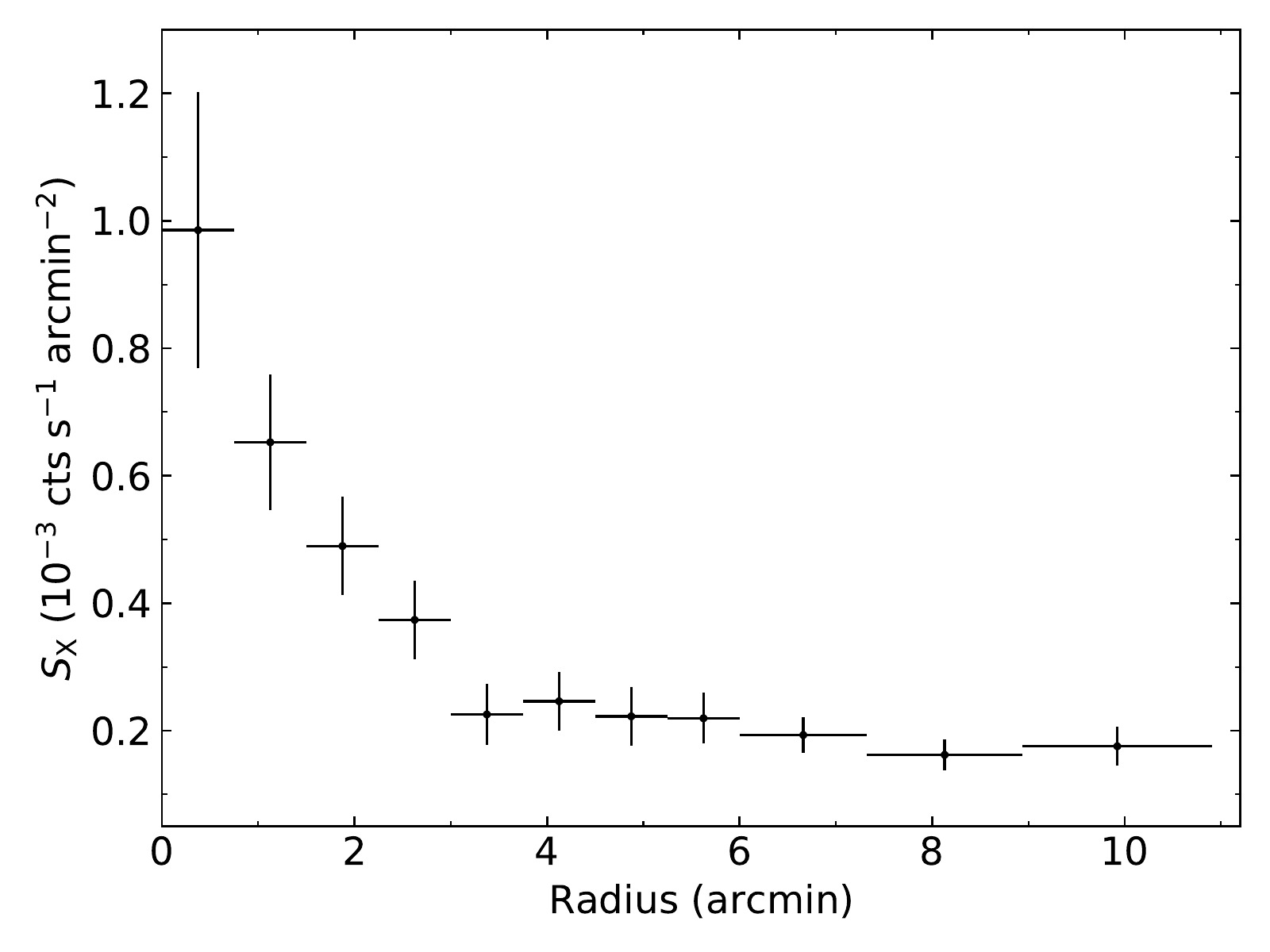} 
\end{center}
\vspace{-0.5cm}
\caption{\textit{XMM--Newton} surface brightness of the background group LOS9, showing that the group emission declines rapidly with increasing radius and reaches the background level at around 3 arcmin from its centre.}
\label{fig: sb_LOS9}
\end{figure}


We derived the surface brightness profile for LOS9 in the soft band by extracting counts in concentric annuli centred on the system, avoiding the regions to the north and south of the group where the two clusters are located. The \textit{XMM--Newton} surface brightness of the background system LOS9 is shown in Fig. \ref{fig: sb_LOS9}. The figure shows that the surface brightness of LOS9 declines rapidly with increasing radius and reaches the background level at 3 arcmin from its centre, implying that an exclusion region of radius 3 arcmin is sufficient to exclude the emission coming from the LOS9 system. We, therefore, exclude the LOS9 group with a 3 arcmin radius circular region. We fitted the LOS9 group’s surface brightness profile with a single $\beta$-model to estimate the contribution of its outskirts to the filament, and we find it to be less than 1 per cent.


Furthermore, we examined the surface brightness profile along the bridge region by fitting it to a single $\beta$-model with the emission of LOS9 included. By fixing the centre of the bridge to the position of the LOS9 centre \citep{Sohn2019the} and fitting for the shape parameters, we find $\beta=0.86_{-0.15}^{+0.19}$ and $r_c=8.87_{-2.13}^{+2.19}$ arcmin, corresponding to $1.91_{-0.46}^{+0.47}$ Mpc at the redshift of the group. These values are significantly larger than those expected for a typical galaxy group \citep[e.g.][]{Helsdon2000intragroup,Dai2010baryon}. For example, \citet{Dai2010baryon} found an average core radius of 290 kpc for a sample of galaxy groups, which is about 7 times smaller than that obtained for LOS9 when fitting for the bridge. We obtain similar values for the shape parameters when the centre of the bridge is left as a free parameter. The estimated centre of the bridge in this case is about 1.5 arcmin away from the position of the LOS9 centre. These findings provide further support to our conclusion that the excess emission seen in the bridge does not originate from LOS9.

\subsection{Spectral analysis}
To examine the spectral properties of the bridge region, we use the \textit{Suzaku} data and extract 2 spectra, one containing the brightest part of the bridge (white dashed circle, labelled `on-filament’ in right-hand panel of Fig. \ref{fig: A2029/2033}) and one adjacent to it (yellow dashed circle, labelled `off-filament’). For each extracted spectrum, ARFS were created using \textsc{xissimarfgen}, RMFs were creates using \textsc{xisrmfgen}, and NXB spectra were obtained using \textsc{xisnxbgen}.

We follow the technique used in \cite{Werner2008detection} when studying the filament in the Abell 222/223 system, where we use the off-filament region (yellow dashed circle) to model emission from the overlap of the two clusters, and incorporate this into the spectral model for the on-filament region, allowing us to fit the spectrum of the excess emission.

We exclude the LOS9 background group ($z \sim 0.223$, identified in \citealt{Sohn2019the}) from the analysis. 
Both of our circle extraction regions (on and off filament) are equidistant from the background group LOS9, so any possible residual background emission from LOS9 should be the same in each region, and will therefore be incorporated into the background modelling.

This method provides a more accurate modelling of the cluster contribution to the spectrum than the method used in the original \textit{Suzaku} analysis in \cite{Walker2012_A2029}, in which the cluster outskirts contribution was modelled using the Abell 2029 emission at the same radial distance but from the southern direction. \cite{Walker2012_A2029} was also limited by the fact that the only X-ray observation of Abell 2033 at the time was the \textit{ROSAT} PSPC observation centred of Abell 2029, which placed Abell 2033 in the outer regions of the PSPC field of view, where the dramatic increase in the \textit{ROSAT} PSF (the half power diameter rises higher than 5 arcmin, \citealt{Boese2000}) meant that its shape was distorted, making it impossible to accurately measure its shape. The \textit{Chandra} and \textit{XMM--Newton} observations of Abell 2033 taken since 2012 allow us to far more accurately measure its expected surface brightness contribution in the bridge region.

As in \cite{Walker2012_A2029}, we use a larger background region (magenta dashed box) to model the X-ray background, as well as RASS data in an annulus reaching from 1--2 degrees around the cluster, which we divide into 4 sectors to find the range of the background fitting components. As in \cite{Walker2012_A2029} and \cite{Vikhlinin2005}, we model the soft galactic foreground with 2 \textit{apec} components at 0.53 keV and 0.20 keV. The cosmic X-ray background (CXB) is modelled as an absorbed power law of index 1.4, whose normalization is determined by integrating the known cumulative flux distribution of point sources (using the double power law model of \citealt{Moretti2003}) down to the threshold flux to which point sources are resolved. The congruent \textit{XMM--Newton} data allow point sources to be resolved down to a threshold flux of $4 \times 10^{-15}$ erg s$^{-1}$ cm$^{-2}$. The contribution from the point sources resolved in the \textit{XMM--Newton} observations are modelled using absorbed power law and added into the background model. We use a column density of $3.26\times10^{20}$ cm$^{-3}$ from \cite{kalberla2005leiden}. For consistency with \citet{Werner2008detection}, we use the same abundance table as them \citep{Lodders2003}.

We fit the off and on-filament regions simultaneously to propagate the uncertainty in the both fits. We perform 10,000 iterations of the fit, each time varying the parameters of the background model through their expected variances, as well as the NXB level (which \cite{Tawa2008} has found to have a systematic uncertainty of $\pm 3$ per cent).

The best fit spectrum is shown in Fig. \ref{fig: SuzakuSpectrum}. We find that the best fit temperature for the excess emission is $1.4^{+0.7}_{-0.5}$ keV, fitting with an absorbed \textit{apec} component. Similar to \citet{Werner2008detection}, we are unable to constrain the metal abundance, so we explore the effect of changing the assumed metal abundance in the spectral modelling through the range 0.0--0.3 Z$_{\odot}$. The systematic uncertainty in the background modelling and in the metal abundance has been propagated through into these quoted error bars on the temperature. The temperature is significantly lower than is expected from the outskirts of Abell 2029 or Abell 2033 near their virial radii (the best fit temperature to the off-filament region modelling the cluster outskirts contribution is 5.1 keV). This is in reasonable agreement with the temperature found in \cite{Werner2008detection} for the gas in the bridge between Abell 222/223.

\section{Discussion}
\label{sec: discussion}

\subsection{Bridge properties}
\label{sec: properties}

To measure the gas density of this excess emission that we see along the bridge connecting Abell 2029 to Abell 2033, we modelled the contributions coming from the two clusters and subtracted their contributions from the bridge region (see Section \ref{sec: SB}). If the observed difference in redshift is only due to the Hubble flow (i.e. there are no peculiar velocity effects), the line-of-sight distance between the two clusters is 20 Mpc. If we assume a cylindrical geometry for the filament, and use our best fit radius (from the mesa model for the bridge) of $5.57_{-0.38}^{+0.39}$ arcmin ($\approx 0.51_{-0.04}^{+0.04}$ Mpc) and a length of 20 Mpc along the line of sight, the derived emission measure for the gas associated with the filament yields an average gas density of $3.7^{+1.0}_{-0.7} \times 10^{-5}$ cm$^{-3}$. These error bars incorporate all of the uncertainties in the background modelling, and also the systematic uncertainty in the metal abundance, which we assume to be in the range 0.0--0.3 Z$_{\odot}$. This corresponds roughly to 160 times the mean baryon density of the Universe, agreeing with that predicted for the densest part of the WHIM \citep[e.g.][]{Bregman2007}. If we assume a constant gas density and a metal abundance of 0.2 Z$_{\odot}$, we then obtain a gas mass of $1.7 \times 10^{13}$ M$_{\odot}$. The total mass of the filament would be $1.1 \times 10^{14}$ M$_{\odot}$, assuming a baryon mass fraction of 0.16. Weak lensing observations of the Abell 222/223 filament by \citet{Dietrich2012Natur} have found a lower gas mass fraction of 0.09. If we assume this value, then the total mass of the filament is $1.9 \times 10^{14}$ M$_{\odot}$. 

The inferred physical properties of the gas in the bridge between Abell 2029 and Abell 2033 are in good agreement with those derived for the filamentary gas between Abell 222 and Abell 223 \citep{Werner2008detection}. Assuming the line-of-sight distance between the two clusters is 15 Mpc, these authors obtained an average gas density of $3.4 \times 10^{-5}$ cm$^{-3}$ (assuming $Z=0.2$ Z$_{\odot}$), which corresponds roughly to 150 times the mean baryon density of the Universe. 
They found a gas temperature of $0.91 \pm 0.25$ keV along the bridge connecting Abell 222 to Abell 223, which is consistent with our best fit temperature ($1.4_{-0.5}^{+0.7}$ keV) for the excess emission between Abell 2029 and Abell 2033. Furthermore, the estimated diameter of the filament from our best model fit to the X-ray data is $1.0 \pm 0.1$ Mpc (see Table \ref{table: best_fitting_parameters}), and this measurement is similar to the diameter of the observed filament in the Abell 222/223 system \citep[1.2 Mpc;][]{Werner2008detection}.    

Cosmological simulations \citep[e.g.][]{Dolag2006Simulating,Martizzi2019Baryons} predicted that the WHIM in filaments that connect clusters of galaxies have a density in approximate range of $5 \times 10^{-6}$ cm$^{-3}$ up to  $10^{-4}$ cm$^{-3}$ in the local Universe. Simulations also found that the hottest part of the WHIM in filaments can have a temperature of around 1 keV. Our density and temperature measurements of the gas in the bridge between Abell 2029 and Abell 2033 are similar to those expected for the WHIM, and might suggest that we are detecting the densest and hottest parts of the WHIM. However, simulations \citep[e.g.][]{vazza2013properties} also predicted a higher level of gas clumping along large-scale filaments. It is therefore possible that part of the observed excess cool emission is the result of unresolved gas clumps. Using the Voronoi tessellation technique, and finding the median surface brightness rather than the mean, we have corrected our results for the effect of gas clumping, following e.g. \citet{eckert2015gas} and \citet{Mirakhor2020complete,Mirakhor2021Virgo}. The spatial scale of the tessellated regions is around $30 \times 30$ kpc$^2$, so we have been able to probe gas clumping down to this spatial scale in the bridge region, but it is still possible that a small part of the observed excess emission in the direction that connects the two clusters is due to unresolved cool gas clumps.  

The inferred density of the gas in the bridge between Abell 2029 and Abell 2033 is determined assuming that the redshift difference between the two clusters is only due to the Hubble flow. If, instead, we assume that the observed difference in redshift is entirely due to peculiar velocity effects, and both clusters would be at the same distance. In this case, the outskirts of Abell 2029 and Abell 2033 would then partially overlap, and the clusters would be in a pre-merger phase before they merge with each other. The gas in the interaction region would have been shock heated due to accretion, and the temperature would be significantly higher than 1 keV. However, this is not the case for the gas associated with the Abell 2029/2033 filament. This suggests that the observed redshift difference between the two clusters is mainly attributed to the Hubble flow, not to peculiar velocities.

\begin{table*}
\begin{minipage}{135mm}
    \centering
    \caption{Statistics of model comparison.}
    \begin{tabular}{ccccccc}
   \hline
    Model & Reduced $\chi^2$ & \multicolumn{3}{c}{Likelihood ratio} & \multicolumn{2}{c}{AIC}   \\
    \cline{3-7}
             &    &  $LR$  & $p$-value  & $\sigma$   &  $\Delta_i$  &   $w_i$                \\           
        \hline
Two double-$\beta$ & 1.40 & --- & ---  & --- & 62.4 & $2.7 \times 10^{-14}$ \\ 
Two double-$\beta$ + single-$\beta$ & 0.97 & 49.9  & $8.2 \times 10^{-11}$  & 6.5 & 6.4 & 0.04 \\ 
Two double-$\beta$ + mesa & 0.91 & 56.4 & $3.5 \times 10^{-12}$ & 7.0 & 0 & 0.96  \\ 
  \hline
    \end{tabular}
    \label{table: model_comparison}
\end{minipage}
\end{table*}

\subsection{Model comparison}
\label{sec: comparison}
\begin{figure}

\begin{center}

\includegraphics[width=1.0\columnwidth]{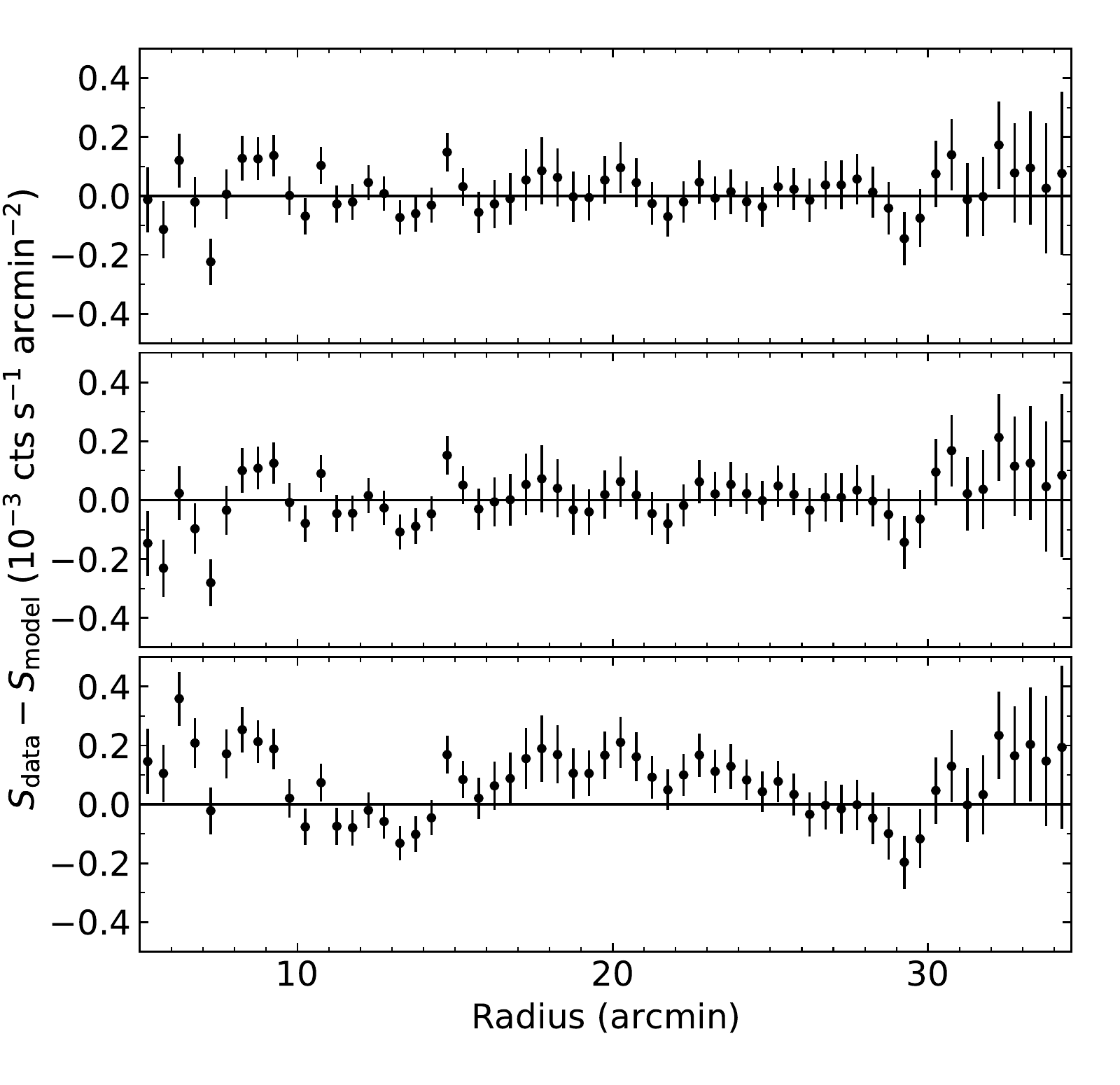} 
\end{center}
\vspace{-0.5cm}
\caption{Residuals of the surface brightness along the bridge region after subtracting the best-fitting models: two-double $\beta \,+$ mesa (\textit{top}), two-double $\beta \,+$ single $\beta$ (\textit{middle}), and two-double $\beta$ (\textit{bottom}). A large residual at the bridge region is apparent when no model for the bridge is considered. The $x$-axis is the distance from the centre of Abell 2029. }
\label{fig: residual}
\end{figure}

As inferred by the reduced $\chi^2$ (see Section \ref{sec: SB} and Table \ref{table: model_comparison}), the models that include a component for the bridge provide a better fit to the data than the model that only includes components for Abell 2029 and Abell 2033 with no component for the bridge. In Fig. \ref{fig: residual}, we compare the residuals of the surface brightness along the bridge region after subtracting the best-fitting models. A large residual at the bridge region is apparent when no model for the bridge is considered. 

To quantify the significance of the bridge component, we test the null hypothesis, i.e. there is no filament between Abell 2029 and Abell 2033, by employing the likelihood-ratio statistic, as is also done in \citet{Hincks2021high}. The likelihood-ratio test can be defined as
\begin{equation}
    LR = -2 \log \frac{\mathcal{L}_{1}^{\rm{max}}}{\mathcal{L}_{2}^{\rm{max}}}, 
    \label{eq: LR}
\end{equation}
where $\mathcal{L}_{1}^{\rm{max}}$ and $\mathcal{L}_{2}^{\rm{max}}$ are the values of the maximum likelihood for the null and alternative hypotheses, respectively. This test can only be applied between two nested models, that is the null hypothesis model with fewer parameters is a special case of the alternative hypothesis model. 

Here, we use the likelihood-ratio test to determine the null hypothesis of the two-double $\beta$-model with ten parameters that is nested in the alternative hypothesis models (the models that include a component for the bridge). When we modelled the surface brightness along the filament direction by adding a "mesa-like" function and a single $\beta$-model to the two-double $\beta$-model, it is found  $LR$ equal to 56.4 and 49.9, respectively. The corresponding $p$-values for the null hypothesis are $3.5 \times 10^{-12}$ and $8.2 \times 10^{-11}$. In other words, a component for the bridge is statistically preferred at about $7.0\sigma$ when modelled with a "mesa-like" function and about $6.5\sigma$ when modelled with a single $\beta$-model. Therefore, based on assumption that Abell 2029 and Abell 2033 can be modelled by two double-$\beta$ components, we find a significant excess of X-ray emission between the two clusters that cannot be explained by the overlap of the clusters. This finding agrees well with that reported for the Abell 399/401 system \citep{Hincks2021high}.

To compare the different models for the bridge itself, we cannot employ the the likelihood-ratio statistic, given that the models are not all nested within each other. However, as is done in \citet{Hincks2021high}, we compute the Akaike Information Criterion \citep[AIC;][]{Akaike1974}:
\begin{equation}
    {\rm{AIC}} = -2 \log \mathcal{L^{\rm{max}}} + 2K, 
    \label{eq: AIC}
\end{equation}
where $K$ is the degrees of freedom (the number of data points minus the number of parameters).


The preferred model is the one with the lowest AIC value. To allow for appropriate comparison of models, it is essential to transform AIC of model $i$ to $\Delta_i = {\rm{AIC}}_i - {\rm{AIC_{min}}}$, where ${\rm{AIC_{min}}}$ is the minimum AIC value. This transformation makes the preferred model to have $\Delta_i = 0$, while makes other models to have positive values. Generally, models with  $\Delta_i \leq 2$ are considered to have "substantial support", those with $4 \leq \Delta_i \leq 7$ have "considerably less support", and models with $\Delta_i > 10$ have "essentially no support" \citep{burnham2004multimodel}. We also calculate the Akaike weights from $\Delta_i$ values as
\begin{equation}
    w_i = \frac{\exp(-\Delta_i/2)}{\sum \exp(-\Delta_r/2)}, 
    \label{eq: w_i}
\end{equation}
where the denominator is the sum of the likelihoods of all models. The Akaike weights, $w_i$, can be interpreted approximately as the probability that model $i$ is the best model, given the data and the set of candidate models \citep[e.g.][]{burnham2004multimodel}.

Table \ref{table: model_comparison} gives statistical information for our model comparison. We find the model that consists of two double-$\beta$ components with a "mesa-like" function has the lowest AIC value, indicating that it is the preferred model. The model that only includes components for the two clusters with no component for the bridge provides a significantly worse fit compared to those that include a bridge component. We also find the model that includes a single-$\beta$ component to fit for the bridge is statistically disfavoured compared to that use a "mesa-like" function. This suggests that the bridge is better represented by a flat, "mesa-like" function than models that peaks more strongly at the centre, such as a single $\beta$-model.

\section{Summary}
\label{sec: summary}
We have explored \textit{XMM--Newton}, \textit{Chandra}, and \textit{Suzaku} data of the Abell 2029/2033 system. After removing emission from background groups, and modelling the surface brightness distribution of the two clusters, we find there to be a significant excess in X-ray emission joining them. This excess in emission cannot be explained by just the overlap of the two clusters. By fitting models to the excess emission, we find that is best modelled with a flat "mesa" type plateau model, with a diameter of 1 Mpc. This "mesa" model provides a better fit than using a $\beta$ model to model the excess emission, and using a $\beta$ model yields a extremely large core radius of 1.1 Mpc. Using the \textit{Suzaku} data to perform a spectral analysis, the temperature of the bridge is $1.4_{-0.5}^{+0.7}$ keV, significantly lower than the outskirts temperatures of Abell 2029 and Abell 2033 in this region ($\sim 5$ keV). The properties of this bridge are, therefore, consistent with a filament with a width of 1 Mpc and a length of 20 Mpc, similar to what has been found in the Abell 222/223 system.

\section*{Acknowledgements}
We thank the referee for helpful comments that improved the paper. This work is based on observations obtained with \textit{Suzaku}, a joint JAXA and NASA mission, and \textit{XMM--Newton}, an ESA science mission with instruments and contributions directly funded by ESA Member States and NASA. This work is also based on observations obtained with the \textit{Chandra} observatory, a NASA mission. 

\section*{Data Availability}
The \textit{Suzaku} data used in the paper are publicly available from the High Energy Astrophysics Science Archive
Research Centre (HEASARC). The \textit{XMM--Newton} Science Archive (XSA) stores the archival data used in this paper, from which the data are publicly available for download. The \textit{XMM} data were processed using the \textit{XMM--Newton} Science Analysis System (\textsc{sas}). The \textit{Chandra} Data Archive stores the data used in this paper. The \textit{Chandra} data were processed using the \textit{Chandra} Interactive Analysis of Observations (\textsc{ciao}) software. The software packages \textsc{heasoft} and \textsc{xspec} were used, and these can be downloaded from the \textsc{heasarc} software web-page. Analysis and figures were produced using \textsc{python} version 3.8.



\bibliographystyle{mnras}
\bibliography{A2029_A2033} 




\appendix


\bsp	
\label{lastpage}
\end{document}